\documentclass[twocolumn,showpacs,preprintnumbers,amsmath,amssymb,superscriptaddress,floatfix,nofootinbib]{revtex4}
\usepackage{graphicx}
\usepackage{amsmath}
\usepackage{amsfonts}
\usepackage{amssymb}%

\newcommand{\Slash}[1]{\ooalign{\hfil/\hfil\crcr$#1$}}

\begin{document}

\title{The near threshold $\pi^- p \to \eta n$ reaction in an effective Lagrangian approach}

\author{Qi-Fang L\"{u}} \affiliation{Department of Physics, Zhengzhou University, Zhengzhou, Henan 450001, China}
\author{Xiao-Hao Liu} \affiliation{Department of Physics, Zhengzhou University, Zhengzhou, Henan 450001, China}

\author{Ju-Jun Xie}~\email{xiejujun@impcas.ac.cn}
\affiliation{Institute of modern physics, Chinese Academy of
Sciences, Lanzhou 730000, China} \affiliation{Research Center for
Hadron and CSR Physics, Institute of Modern Physics of CAS and
Lanzhou University, Lanzhou 730000, China} \affiliation{State Key
Laboratory of Theoretical Physics, Institute of Theoretical Physics,
Chinese Academy of Sciences, Beijing 100190, China}

\author{De-Min Li}~\email{lidm@zzu.edu.cn} \affiliation{Department of Physics, Zhengzhou University, Zhengzhou, Henan 450001, China}

\begin{abstract}

The near threshold $\pi^- p \to \eta n$ reaction is studied within
an effective Lagrangian approach and the isobar model. By
considering the contributions from $s$- and $u$-channel nucleon pole
and $N^{*}(1535)$ resonance, the total and differential cross
sections of the $\pi^- p \to \eta n$ reaction near threshold are
calculated. Our theoretical results can fairly reproduce the current
experimental data. It is also shown that while the center-of-mass
energy lies in the range from the reaction threshold up to $1.65$
GeV, $s$-channel $N^{*}(1535)$ resonance plays the dominant role.
The effect from nucleon pole is found to be small but the
interference terms between the $N^*(1535)$ resonance and the nucleon
pole are significant. The contributions from $t$-channel processes
are negligible in the present calculation.

\end{abstract}

 \maketitle

\section{Introduction}

The study of  the nucleon and its excited states is an interesting
topic in hadron physics. In classical constituent quark models
(CQM), proton is described as a three-quark ($uud$) state. This
picture is very successful in explaining the properties of the
spatial ground states, but not for the case of the excited states.
For example, the $N^{*}(1535)$ resonance, four-star state in the
Particle Data Review Book (PDG)~\cite{pdg2012}, with spin-parity
$J^{p} = 1/2^-$, is expected to be the lowest $L=1$ orbital excited
nucleon state~\cite{Zou:2009zz,Capstick:1986bm} according to the
CQM. However, the $N^*(1535)$ resonance is heavier than the spatial
excited nucleon state, $N^*(1440)$ ($J^{p} = 1/2^+$). This is the
long standing inverse mass problem for the low-lying excited nucleon
states. Furthermore, it is well known that $N^{*}(1535)$ resonance
couples strongly to the final states with strangeness, such as,
$\eta N$
channel~\cite{pdg2012,Shklyar:2012js,He:2009zzr,Zhong:2007fx}, $K
\Lambda$ channel~\cite{liuplk,liuplk1,liuplk2}, $\eta'
N$~\cite{Cao:2008st,Zhong:2011ht}, and $\phi N$
channel~\cite{xiephi,Cao:2009ea,daiphi}, which implies a
considerable amount of $s\bar{s}$ component in the $N^*(1535)$ wave
function~\cite{Zou:2009zz,An:2009zza,An:2010zz}.

On the other hand, the $\pi^{-}p \to \eta n$ reaction is of
particular interest in studying the structure of the $N^*(1535)$
resonance, of which the properties still bare a lot of
controversies. Since there are no isospin-$3/2$ $\Delta$ baryons
contributing here, this reaction gives us a rather clean platform to
study the isospin $1/2$ nucleon resonances, especially for studying
the $N^*(1535)$ resonance because it couples strongly to the $\eta
N$ channel. This reaction has been theoretically studied by using a
chiral quark-model approach in Refs.~\cite{He:2009zzr,Zhong:2007fx},
and also in Ref.~\cite{Shklyar:2012js} with an updated
coupled-channel method. They all found that the contributions from
the $N^*(1535)$ resonance dominate the reaction near threshold.
However, the present theoretical calculations are still far from
being as accurate as the experiment~\cite{Zhong:2007fx}. Thus, more
theoretical studies are welcome.

Along this line, with the near threshold experimental
data~\cite{Prakhov:2005qb,Bulos:1970zk,Deinet:1969cd,Brown:1979ii},
we reanalysis the $\pi^{-}p \to \eta n$ reaction from the production
threshold to the center-of-mass energy $W \simeq 1.65$ GeV by using
the effective Lagrangian approach and the isobar model. We payed
especial attention to the role of the $N^*(1535)$ resonance, while
the contribution of nucleon pole is also considered in the present
calculation, and we find that the interference terms between the
$N^*(1535)$ resonance and the nucleon pole are significant.
Moreover, in those previous works, they all take a constant total
decay width, $150$ MeV, for $N^*(1535)$ resonance. In this work,
both the energy-dependent total width and the constant total width
for the $N^*(1535)$ resonance are used.

This paper is organized as follows. In Sect.~\ref{sec:formalism}, we
shall discuss the formalism and the main ingredients of the model.
The numerical results and discussions are presented in
Sect.~\ref{sec:results}. Finally, a short summary is given in the
last section.

\section{Formalism and ingredients} \label{sec:formalism}

\begin{figure}[htbp]
\begin{center}
\includegraphics[scale=0.45]{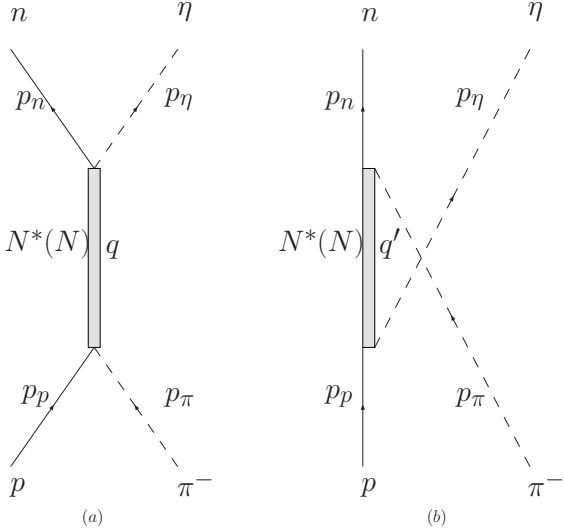}
\caption{Feynman diagrams for $\pi^{-}p \to \eta n$ reaction. In
these diagrams, the contributions from $s$-channel and $u$-channel
nucleon pole and $N^*(1535)$ resonance are considered. We also show
the definitions of the kinematical ($p_{\pi}, p_{p}, p_{\eta}, p_n,
q, q'$) those we use in our calculation.} \label{Fig:feyndgm}
\end{center}
\end{figure}

As shown in
Refs~\cite{liuplk,liuplk1,liuplk2,xiephi,Tsushima:2000hs,Sibirtsev:2005mv,Shyam:1999nm,Xie:2007vs,xie1:2007vs,gaopzkn,gaopzkn1,Liu:2011sw,Liu1:2011sw},
the combination of the effective Lagrangian approach and the isobar
model is a good method to study the hadron resonances production in
the $\pi N$, $NN$, and $\bar{K}N$ scattering. In this work, we will
use this approach to study the near threshold $\pi^{-}p \to\eta n$
reaction. The basic tree level Feynman diagrams for $\pi^{-}p \to
\eta n$ reaction are depicted in Fig.~\ref{Fig:feyndgm}, where
contributions from the $s$-channel and $u$-channel nucleon pole and
$N^{*}(1535)$ ($\equiv N^*$) resonance are considered. The
contributions from the $t$-channel $a_0(980)$ exchange are ignored,
because the information of $a_0 NN$ vertex is scarce and the mass of
$a_0(980)$ is heavy, which will suppress its contribution.

To compute those terms shown in Fig.~\ref{Fig:feyndgm}, we take the
effective Lagrangian densities as
\begin{eqnarray}
{\mathcal L}_{\pi N N}  &=& i g_{\pi N N} \bar{N} \gamma_5 \vec\tau
\cdot \vec\pi N, \label{pinn}  \\
{\mathcal L}_{\eta N N}  &=& i g_{\eta N N} \bar{N} \gamma_5
\eta N, \label{etann}  \\
\mathcal{L}_{\pi N N^{*}} &=& -ig_{\pi N N^{*}} \bar{N} \vec\tau \cdot \vec\pi N^* + h.c., \label{pinnstar} \\
\mathcal{L}_{\eta N N^{*}}  &=& -ig_{\eta N N^{*}}  \bar{N} \eta N^*
+ h.c., \label{pinnstar}
\end{eqnarray}
with $g^2_{\pi NN}/4\pi =14.4 $ and $g^2_{\eta NN}/4\pi =0.4$ as
used in Ref.~\cite{xiephi}. The values of coupling constants $g_{\pi
N N^{*}}$ and $g_{\eta N N^{*}}$ can be determined from the
$N^*(1535)$ partial decay widths,
\begin{eqnarray}
\Gamma_{N^{*} \to N\pi} & = &
\dfrac{3g_{N^{*}N\pi}^{2}(m_N+E_N^{\pi})p_{\pi}^{\rm c.m.}}{4\pi
M_{N^{*}}}, \\
\Gamma_{N^{*} \to N\eta} & = & \frac{g_{N^{*}N\eta}^{2}(m_N +
E_N^{\eta})p_{\eta}^{\rm c.m.}}{4\pi M_{N^{*}}},
\end{eqnarray}
where $m_N, m_{\pi}, m_{\eta}$, and $M_{N^*}$ are the masses of
proton, $\pi$ meson, $\eta$ meson, and $N^*(1535)$ resonance,
respectively. The $p_{\pi/\eta}^{\rm c.m.}$ is the magnitude of the
3-momentum of $\pi/\eta$ meson that was measured in the $N^*(1535)$
resonance rest frame, and $E_N^{\pi/\eta}$ is the energy of proton
in the $N\pi$ or $N\eta$ decay. The $p_{\pi/\eta}^{\rm c.m.}$ and
$E_N^{\pi/\eta}$ have the following forms,
\begin{eqnarray}
p_{\pi / \eta}^{\rm c.m.} &=& \frac{\lambda^{1/2}(M_{N^*},m_N,m_{\pi/\eta})}{2M_{N^{*}}}, \\
E_N^{\pi / \eta} &=& \sqrt{(p_{\pi / \eta}^{\rm c.m.})^{2}+m_N^{2}},
\end{eqnarray}
where $\lambda$ is the K\"allen function with $\lambda(x,y,z) =
(x-y-z)^2-4yz$.

With experimental mass ($1535$ MeV), total decay width ($150$ MeV),
and the branch ratios of $N^{*}$(1535) resonance quoted in the
PDG~\cite{pdg2012}, we obtain $g^2_{\pi NN^*}/4\pi =0.037$ and
$g^2_{\eta NN^*}/4\pi = 0.28$.

Next, we pay attention to the total scattering amplitude
$\mathcal{M}$ of $\pi^{-}p \to \eta n$ reaction,
\begin{eqnarray}
\mathcal{M} &=& \mathcal{M}_s + \mathcal{M}_u \nonumber
\\
&=& \mathcal{M}^N_s + \mathcal{M}^{N^*}_s + \mathcal{M}^N_u +
\mathcal{M}^{N^*}_u.
\end{eqnarray}

Each amplitude can be obtained straightforwardly with the effective
Lagrangian densities shown above. Here we give explicitly the
amplitudes, $\mathcal{M}^{N}_s$ and $\mathcal{M}^{N^*}_s$, as an
example,
\begin{eqnarray}
\mathcal{M}^{N}_s &=& - \sqrt{2}g_{NN\pi}g_{NN\eta}F_{N}(s) \times
\nonumber
\\ && \bar{u}(p_n,s_n) \gamma_5 G_{N}(s) \gamma_5 u(p_p,s_p), \\
\mathcal{M}^{N^*}_s &=&
\sqrt{2}g_{N^{*}N\pi}g_{N^{*}N\eta}F_{N^{*}}(s) \times \nonumber
\\ && \bar{u}(p_n,s_n)G_{N^{*}}(s)u(p_p,s_p),
\end{eqnarray}
with $s = W^2= (p_p + p_{\pi})^2$, the invariant mass square of the
$\pi^- p$ system. The $F_N(s)$ and $G_{N}(s)$ [ $F_{N^*}(s)$ and
$G_{N^{*}}(s)$ ] are respectively the form factor and propagator for
the nucleon pole[( $N^*(1535)$ resonance ].

The form factors for nucleon pole and $N^{*}$(1535) resonance,
$F_N(s)$ and $F_{N^{*}}(s)$, are introduced to describe the
off-shell properties of the amplitude, and we choose the forms of
them as,
\begin{eqnarray}
F_{N}(s) &=& \frac{\Lambda_{N} ^{4}}{\Lambda_{N}
^{4}+(s-m^2_{N})^{2}}, \\
F_{N^{*}}(s) &=& \frac{\Lambda_{N^*} ^{4}}{\Lambda_{N^*}
^{4}+(s-M^2_{N^{*}})^{2}},
\end{eqnarray}
with $\Lambda_N = 0.6$ GeV and $\Lambda_{N^*} = 2.0$ GeV as used in
Ref.~\cite{xiephi} for the $\pi^{-}p \to\phi n$ reaction.

For the propagators $G_{N}(s)$ and $G_{N^{*}}(s)$ of the nucleon
pole and the $N^{*}$(1535) resonance in the $s-$channel, we take
them as~\cite{liangjpg28},
\begin{eqnarray}
G_{N}(s) &=& \frac{i(\Slash{q} + m_{N})}{s - m_{N}^{2}}, \\
G_{N^{*}}(s) &=& \frac{i(\Slash{q}+M_{N^{*}})}{s - M_{N^{*}}^{2} +
iM_{N^{*}}\Gamma_{N^{*}}(s)},
\end{eqnarray}
where $\Gamma_{N^{*}}(s)$ is the $N^*(1535)$ energy-dependent total
decay width. Since the main decay channels of $N^*(1535)$ resonance
are the $\pi N$ and $\eta N$, we take the commonly used phase space
dependent width for the $N^*(1535)$ resonance
as~\cite{liuplk,liuplk1,liuplk2},
\begin{eqnarray}
\Gamma_{N^{*}}(s) \! = \!\! \Gamma_{N^{*}\to N\pi} \!
\frac{\rho_{\pi N}(s)}{\rho_{\pi  N}(M_{N^{*}}^{2})} \! + \!
\Gamma_{N^{*}\to N\eta} \! \frac{\rho_{\eta N}(s)}{\rho_{\eta
N}(M_{N^{*}}^{2})} \, , \label{gamr-ed}
\end{eqnarray}
where $ \Gamma_{N^{*}\to N\pi} = 75$ MeV and $\Gamma_{N^{*}\to
N\eta} = 75$ MeV are the $N^*(1535)$ partial decay widths, the
$\rho_{\pi N}(s)$ and $\rho_{\eta N}(s)$ are the phase space factors
for $\pi N$ and $\eta N$ final states, respectively, for example,
\begin{eqnarray}
\rho_{\pi N} (s) \!\! = \!\! \frac{\sqrt{(s - (m_N+m_{\pi})^2)(s -
(m_N-m_{\pi})^2)}}{\sqrt{s}}.
\end{eqnarray}

From the scattering amplitudes given above, we can calculate the
total and differential cross sections for $\pi^{-}p \to \eta n$
reaction. At the center of mass (c.m.) frame, the differential cross
section for $\pi^{-}p \to \eta n$ reaction can be expressed as,
\begin{eqnarray}
\frac{d\sigma}{dcos\theta} = \frac{m^2_N}{8\pi s}
\frac{|\vec{p_{\eta}}|}{|\vec{p_{\pi}}|}(\frac{1}{2}\sum_{s_p,s_n}|\mathcal{M}|^{2}),
\label{eq:dcs}
\end{eqnarray}
where $s_p$ and $s_n$ are the spin polarization variables of initial
proton and final neutron, respectively. The $\theta$ is the angle of
the outgoing $\eta$ meson relative to the beam direction in the c.m.
frame, while $\vec{p_{\pi}}$ and $\vec{p_{\eta}}$ are the 3-momentum
of the initial $\pi^-$ and the final $\eta$ mesons, respectively.

\section{Numerical Results and discussion} \label{sec:results}

\begin{figure}[htbp]
\begin{center}
\includegraphics[scale=0.65]{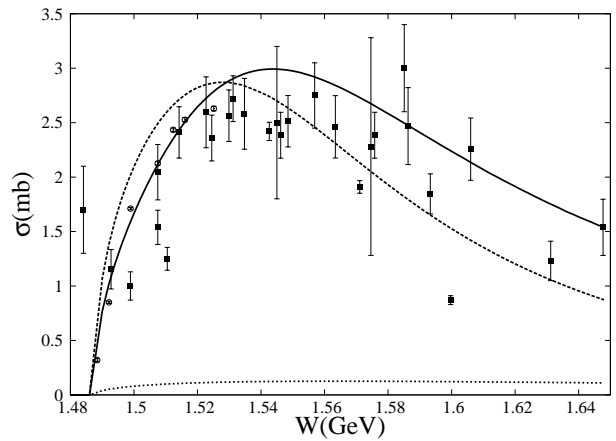}
\caption{Total cross section versus center of mass energy $W$ for
$\pi^{-}p \to n \eta$ reaction with the energy-dependent
$\Gamma_{N^*}(s)$. The dotted and dashed curves stand for the
contributions from the nucleon pole and the $N^*(1535)$ resonance,
respectively, while the solid line stands for the total
contributions. The data are from Ref.~\cite{Prakhov:2005qb} (dot)
and Ref.~\cite{Bulos:1970zk} (square).} \label{Fig:tcsed}
\end{center}
\end{figure}

Firstly, with Eq.~(\ref{eq:dcs}), we calculate the total and
differential cross sections of $\pi^- p \to \eta n$ reaction from
the production threshold up to the center-of-mass energy $W=1.65$
GeV by using the $N^*(1535)$ energy-dependent width as in
Eq.~(\ref{gamr-ed}). The corresponding theoretical results as well
as the experimental data from
Refs.~\cite{Prakhov:2005qb,Bulos:1970zk} are shown in
Fig.~\ref{Fig:tcsed}, where the dotted and dashed curves stand for
the contributions from the nucleon pole and the $N^*(1535)$
resonance, respectively, while the solid line stands for the total
contributions, which can describe well the experimental data with
$\Lambda_N = 0.6$ GeV and $\Lambda_{N^*} = 2.0$ GeV.

From Fig.~\ref{Fig:tcsed}, we can see that the contributions from
$N^*(1535)$ resonance are dominant, but not enough to reproduce the
experimental data, whereas the contributions from the nucleon pole
are minor, but, the interference terms of them are significant. We
also find that the results are sensitive to the cutoff parameter,
$\Lambda_N$, but not to the cutoff parameter, $\Lambda_{N^*}$, which
is because the form factor for $N^*(1535)$ is close to $1$ with the
invariant mass $W$ around $1535$ MeV in the present calculation.

\begin{figure}[htbp]
\begin{center}
\includegraphics[scale=0.65]{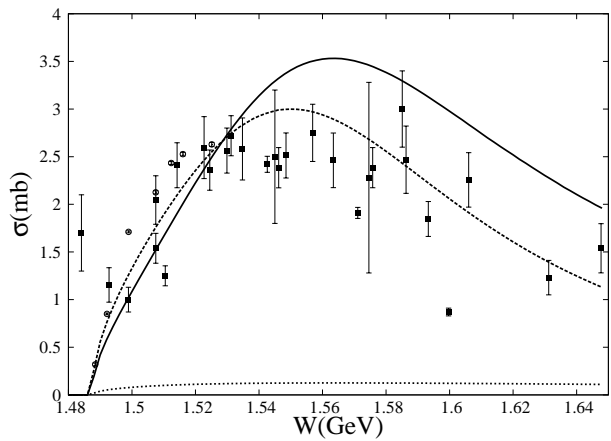}
\caption{As in Fig.~\ref{Fig:tcsed}, but for the case of
$\Gamma_{N^*} = 150$ MeV.} \label{Fig:tcsct}
\end{center}
\end{figure}

On the other hand, we also calculate the total cross section by
using a constant total decay width $\Gamma_{N^*} = 150$ MeV for
$N^*(1535)$ resonance, the results are shown in
Fig.~\ref{Fig:tcsct}. In this case, the contributions from the
$N^*(1535)$ resonance are absolutely dominant, and the experimental
data can be reasonably reproduced by only considering the
$N^*(1535)$ resonance. The contributions from nucleon pole are minor
and the interference terms are significant, which is the same as the
case with the energy-dependent total decay width $\Gamma_{N^*} (s)$.
This is in agreement with the previous
calculations~\cite{Shklyar:2012js,He:2009zzr,Zhong:2007fx}.

From the results, dashed lines in Fig.~\ref{Fig:tcsed} and
Fig.~\ref{Fig:tcsct}, we find that the Breit-Wigner mass of
$N^*(1535)$ resonance will be pushed down if we use the
energy-dependent width, which is similar to the case of
$\Lambda(1405)$ state that we found in Ref.~\cite{xie1405}. This
will have important implications on various model calculations on
the mass of $N^*(1535)$ resonance~\cite{liuplk,liuplk1,liuplk2}.

\begin{figure*}
\begin{center}
\includegraphics[scale=1.3]{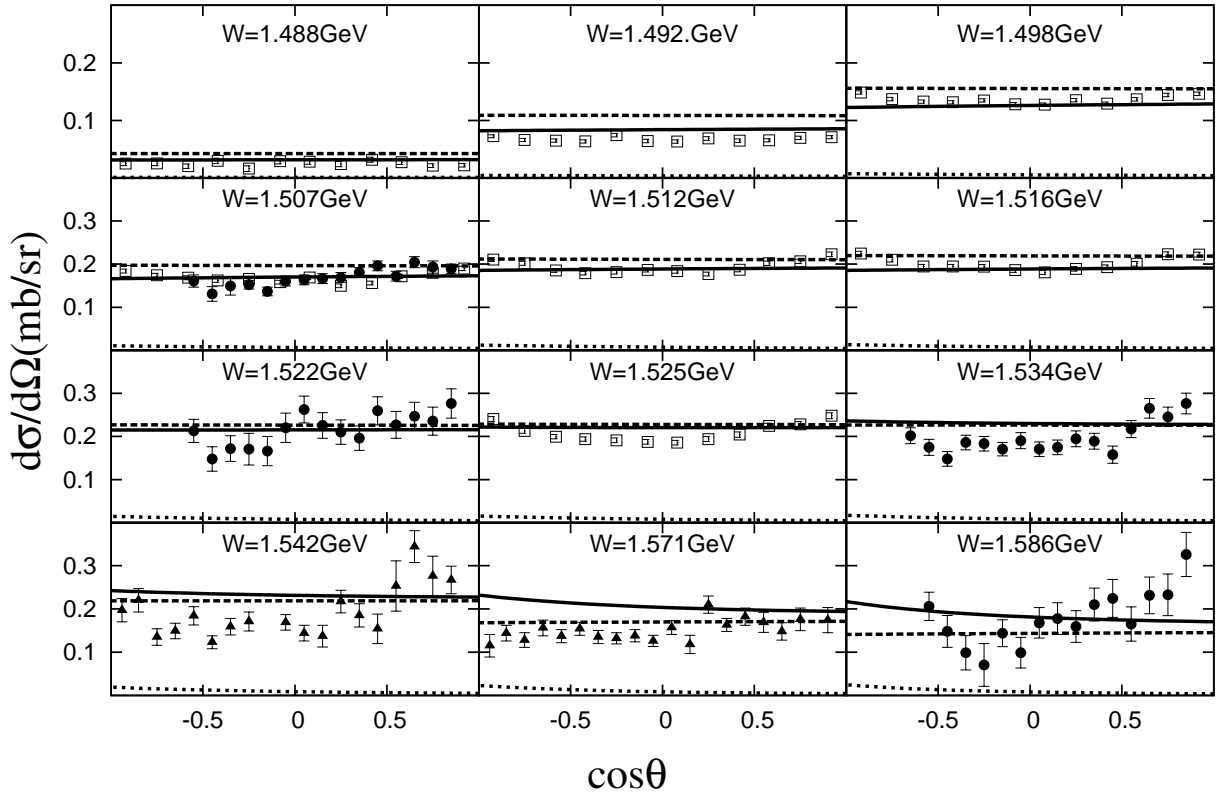}
\caption{Differential cross sections for $\pi^- p \to \eta n$
reaction as a function of ${\rm cos}\theta$ at different
center-of-mass energy, $W$, in the presence of the $N^*(1535)$ total
width being energy-dependent. The dotted and dashed curves stand for
the contributions from the nucleon pole and the $N^*(1535)$
resonance, respectively; the solid line stands for the total
contributions. The experimental data are from
Refs.~\cite{Prakhov:2005qb}(open square),
Ref.~\cite{Deinet:1969cd}(dot), and
Ref.~\cite{Brown:1979ii}(triangle).} \label{Fig:dcsed}
\end{center}
\end{figure*}

Furthermore, the differential cross section of the $\pi^- p \to n
\eta$ reaction with the energy-dependent total decay width
$\Gamma_{N^*} (s)$ for the $N^*(1535)$ resonance~\footnote{We do not
show our theoretical results by using a constant total decay width
since the similar results have been shown in the previous
works~\cite{Shklyar:2012js,He:2009zzr,Zhong:2007fx}.} is calculated
and shown in Fig.~\ref{Fig:dcsed}. We can see that our theoretical
results can reasonably describe the experimental data, especially
for those energy points near reaction threshold thanks to the main
contributions from the $N^*(1535)$ resonance and the significant
interference between the $N^*(1535)$ resonance and the nucleon pole.

\section{Summary}

The $\pi^{-}p \to \eta n$ reaction near-threshold is studied in the
frame of the effective Lagrangian method and the isobar model, which
have been extensively used to deal with hadron collisions. We
calculate the total and differential cross sections for this
reaction by considering the contributions from the $N^*(1535)$
resonance and the nucleon pole. The energy-dependent total width and
the constant total width for the $N^*(1535)$ resonance are used. In
both cases, the contributions from the $N^*(1535)$ resonance are
absolutely dominant. In the case of the $N^*(1535)$ total decay
width being $150$ MeV, the experimental data can be reasonably
accounted for by only considering the N*(1535) resonance, which is
in agreement with the previous
calculations~\cite{He:2009zzr,Zhong:2007fx,Shklyar:2012js}.

From our results, it is shown that the $s$-channel $N^{*}(1535)$
resonance exchange plays the dominated role from the reaction
threshold to the center-of-mass energy $W \simeq 1.65$ GeV. The
effect from nucleon pole is found to be small but the interference
terms between the $N^*(1535)$ resonance and the nucleon pole are
significant. We also find that the Breit-Wigner mass of $N^*(1535)$
resonance will be pushed down if we use the energy-dependent total
width, which will have important implications on various model
calculations on its mass.

\section*{Acknowledgments}

This work is partly supported by the National Natural Science
Foundation of China under grant 11105126.


\begin{thebibliography}{99}
%
\bibitem{pdg2012} J. Beringer \emph{et al.} (Particle Data Group), Phys. Rev. \textbf{D 86}, 010001
(2012).

\bibitem{Zou:2009zz}
  B.~S.~Zou,
  Nucl.\ Phys.\ A {\bf 827}, 333C (2009).  
\bibitem{Capstick:1986bm}
  S.~Capstick and N.~Isgur,
  Phys.\ Rev.\ D {\bf 34}, 2809 (1986).  

\bibitem{Shklyar:2012js}
  V.~Shklyar, H.~Lenske and U.~Mosel,
  Phys.\ Rev.\ C {\bf 87}, 015201 (2013). 
\bibitem{He:2009zzr}
  J.~He and B.~Saghai,
  Chin.\ Phys.\ C {\bf 33}, 1389 (2009).  
  Phys.\ Rev.\ C {\bf 80}, 015207 (2009). 
\bibitem{Zhong:2007fx}
  X.~H.~Zhong, Q.~Zhao, J.~He and B.~Saghai,
  Phys.\ Rev.\ C {\bf 76}, 065205 (2007). 

%
\bibitem{liuplk} B. C. Liu and B. S.~Zou, Phys.\ Rev.\ Lett.\ \textbf{96}, 042002
(2006).

\bibitem{liuplk1} B. C. Liu and B. S.~Zou, Phys.\ Rev.\ Lett.\ \textbf{98}, 039102
(2007).

\bibitem{liuplk2} B. C. Liu and B. S.~Zou, Commun.\ Theor.\ Phys.\ \textbf{46}, 501 (2006).
%
\bibitem{Cao:2008st}
  X.~Cao and X.~G.~Lee,
  Phys.\ Rev.\ C {\bf 78}, 035207 (2008). 
\bibitem{Zhong:2011ht}
  X.~H.~Zhong and Q.~Zhao,
  Phys.\ Rev.\ C {\bf 84}, 065204 (2011). 

\bibitem{xiephi}J. J. Xie, B. S. Zou, and H. C. Chiang, Phys. Rev. C {\bf 77},
015206(2008).

\bibitem{Cao:2009ea}
  X.~Cao, J.~J.~Xie, B.~S.~Zou and H.~S.~Xu,
  Phys.\ Rev.\ C {\bf 80}, 025203 (2009). 
\bibitem{daiphi}J. P. Dai, P. N. Shen, J. J. Xie, and B. S. Zou, Phys. Rev. D {\bf 85},
014011(2012).
\bibitem{An:2009zza}
  C.~S.~An,
  Chin.\ Phys.\ C {\bf 33}, 1393 (2009).  
\bibitem{An:2010zz}
  C.~S.~An and B.~S.~Zou,
  Chin.\ Phys.\ C {\bf 34}, 245 (2010).  

\bibitem{Prakhov:2005qb}
  S.~Prakhov, B.~M.~K.~Nefkens, C.~E.~Allgower, R.~A.~Arndt, V.~Bekrenev, W.~J.~Briscoe, M.~Clajus and J.~R.~Comfort {\it et al.},
  Phys.\ Rev.\ C {\bf 72}, 015203 (2005).  

\bibitem{Bulos:1970zk}
  F.~Bulos, R.~E.~Lanou, A.~E.~Piper, A.~M.~Shapiro, C.~A.~Bordner, A.~E.~Brenner, M.~E.~Law and E.~E.~Ronat {\it et al.},
  Phys.\ Rev.\  {\bf 187}, 1827 (1969).  


\bibitem{Deinet:1969cd}
  W.~Deinet, H.~Mueller, D.~Schmitt, H.~M.~Staudenmaier, S.~Buniatov and E.~Zavattini,
  Nucl.\ Phys.\ B {\bf 11}, 495 (1969).  
\bibitem{Brown:1979ii}
  R.~M.~Brown, A.~G.~Clark, P.~J.~Duke, W.~M.~Evans, R.~J.~Gray, E.~S.~Groves, R.~J.~Ott and H.~R.~Renshall {\it et al.},
  Nucl.\ Phys.\ B {\bf 153}, 89 (1979).  

\bibitem{Tsushima:2000hs}
  K.~Tsushima, A.~Sibirtsev and A.~W.~Thomas,
  Phys.\ Rev.\ {\bf C 62}, 064904 (2000).  

\bibitem{Sibirtsev:2005mv}
  A.~Sibirtsev, J.~Haidenbauer, H.~W.~Hammer and S.~Krewald,
  Eur.\ Phys.\ J.\ {\bf A  27}, 269 (2006). 
\bibitem{Shyam:1999nm}
  R.~Shyam,
  Phys.\ Rev.\ {\bf C 60}, 055213 (1999).  
%

\bibitem{Xie:2007vs}
  J.~J.~Xie and B.~S.~Zou,
  Phys.\ Lett.\ {\bf B 649}, 405 (2007).

\bibitem{xie1:2007vs} J.~J.~Xie, B.~S.~Zou and B.~C.~Liu,
  Chin.\ Phys.\ Lett.\  {\bf 22}, 2215 (2005).
\bibitem{gaopzkn} P. Z. Gao, B. S. Zou and A. Sibirtsev, Nucl. Phys. {\bf A 867}, 41
(2011).
\bibitem{gaopzkn1} P. Z. Gao, J. Shi, and B. S. Zou, Phys.\ Rev.\ {\bf C  86},
025201 (2012). 

\bibitem{Liu:2011sw}
  B.~-C.~Liu and J.~J.~Xie,
  Phys.\ Rev.\ C {\bf 85}, 038201 (2012). 
\bibitem{Liu1:2011sw} B.~-C.~Liu and J.~J.~Xie, Phys.\ Rev.\ C {\bf 86}, 055202 (2012). 
\bibitem{liangjpg28}W. H. Liang, P.N. Shen, J. X. Wang, and B. S. Zou, J. Phys. G {\bf 28}, 333
(2002).


\bibitem{xie1405}J. J. Xie, B. C. Liu, and C. S. An, Phys.
Rev. C {\bf 88}, 015203 (2013).


\end{thebibliography}
\end{document}